\documentclass[12pt]{article}

\usepackage{amsmath,amsfonts}
\usepackage{amssymb}
\usepackage{hhline}
\usepackage{epsfig,cite}
\usepackage{stmaryrd}
\usepackage[usenames,dvips]{color}
\usepackage{fullpage}
\usepackage{verbatim}
   \allowdisplaybreaks
\makeatletter
\@addtoreset{equation}{section}
\makeatother

\newcommand{\be}{\begin{equation}}
\newcommand{\ee}{\end{equation}}
\newcommand{\bea}{\begin{eqnarray}}
\newcommand{\eea}{\end{eqnarray}}

\def\simlt{\stackrel{<}{{}_\sim}}
\def\simgt{\stackrel{>}{{}_\sim}}

\begin{document}

\thispagestyle{empty}

\begin{center}
\hfill UAB-FT-756\\
\hfill EFI-14-14
\begin{center}

\vspace{.5cm}

{\Large\sc Focus Point in the Light Stop Scenario}

\end{center}

\vspace{1.cm}

\textbf{ A.~Delgado$^{\,a}$, 
M.~Quiros$^{\,b}$ and C.~E.~M.~Wagner$^{\,c}$}\\

\vspace{1.cm}
${}^a\!\!$ {\em {Department of Physics, University of Notre Dame\\ Notre Dame, IN 46556, USA}}

\vspace{.1cm}

${}^b\!\!$ {\em {
Instituci\'o Catalana de Recerca i Estudis  
Avan\c{c}ats (ICREA) and\\ Institut de F\'isica d'Altes Energies, Universitat Aut{\`o}noma de Barcelona\\
08193 Bellaterra, Barcelona, Spain}}

\vspace{.1cm}
${}^c$ {\em{Enrico Fermi Institute,  Department of Physics and Kavli Institute for \\ Cosmological Physics,
University of Chicago, Chicago, IL 60637, U.S.A. \\
HEP Division, Argonne National Laboratory, Argonne, IL 60439, U.S.A. \\
}}

\end{center}

\vspace{0.8cm}

\centerline{\bf Abstract}
\vspace{2 mm}
\begin{quote}\small
The recent discovery of a light CP-even Higgs in a region of masses consistent with the predictions of models with low energy supersymmetry  have intensified the discussion of naturalness in these situations.   The
focus point solution alleviates the MSSM fine tuning problem. In a previous work, we showed the general form of the MSSM focus point solution, for different values of the messenger scale and of the ratio of gaugino and scalar masses.  Here we study the possibility of inducing a light stop as a result of the 
renormalization group running from high energies. This scenario is highly predictive and leads to observables that may be constrained by future collider and
flavor physics data.  

 \end{quote}

\vfill

 \newpage
\section{Introduction}

The recent discovery of a Standard Model (SM)-like Higgs boson at the LHC~\cite{ATLAS,CMS}, has renewed the interest in weakly coupled extensions of the Standard Model, in which the effects of new physics decouple rapidly as the new physics scale goes above the weak scale. Such models include the minimal supersymmetric extension of the SM (MSSM)~\cite{reviews}.  This model includes two Higgs doublets and the lightest CP-even Higgs mass is determined in the limit when the rest of the Higgs spectrum is heavy by the neutral gauge boson mass, $M_Z$, the ratio of the two vacuum expectation values $\tan\beta$, and the stop mass spectrum. The measured value of the Higgs mass, $m_h \simeq 125$~GeV,  may be obtained for moderate or large values of $\tan\beta$ and a stop spectrum at the TeV scale, provided the stop mixing parameter $X_t$ is larger than the average stop mass scale. Larger values of the stop mass scale would be necessary for smaller values of the stop mixing parameter. In particular for small values of $X_t$ and $\tan\beta > 5$, the stop mass scale must be of the order of a few tens of TeV~\cite{Giudice:2011cg,Hahn:2013ria,Delgado:2013gza,Draper:2013oza}.  

The physical Higgs mass $m_h$ has a logarithmic dependence on the stop mass scale. However the Higgs mass parameter depends on a quadratic way on this scale. Since the Higgs mass parameter must be of order $m_h$ for a proper minimization of the effective Higgs potential, a fine tuning between the tree-level Higgs mass parameter and the 
radiative corrections must be present. Since the radiative corrections, and therefore the fine tuning, quadratically increase with the stop mass scale, fine tuning arguments lead to the preference of models in which the stops are not heavier than a few TeV and the mixing parameter $X_t$ is sizeable at low energies.  Even in the case of stops at the TeV scale, there is the question on the conditions to obtain a Higgs mass parameter at the weak scale when the stop spectrum is of the order of a few TeV, particularly in the case of large messenger scales, in which the radiative corrections to the Higgs mass parameter are increased due to the logarithmic dependence on the messenger scale. 

In a previous article~\cite{Delgado:2014vha}, we analyzed the general conditions under which the low energy Higgs mass parameter becomes independent of the overall
supersymmetric particle mass scale.  We call these supersymmetric models generalized focus point (FP) scenarios, since a particular case is the focus point~\cite{Chan:1997bi,Feng:1999zg,Feng:2012jfa}, in which the scalar masses are universal at the Grand Unification scale $M_{\rm GUT}$ and the gaugino mass parameters are much smaller than the scalar masses. In these models the radiative corrections are not small, but the dependence of the 
Higgs mass parameter on the boundary value at the messenger scale is appropriately cancelled with the tree-level Higgs mass parameters at the messenger scale. The precise cancellation would, in principle, imply a large fine tuning, since it would imply a precise correlation between the values of the Higgs and stop supersymmetric mass parameters at the messenger scale. The idea behind the focus point is that the necessary correlations occur naturally in specific supersymmetry breaking scenarios, eliminating the need for fine tuning of the mass parameters.

In this article we will analyze the possibility that these correlations could also lead to a  light stop in the spectrum in the generalized focus point scenarios. In 
section~\ref{LightStops} we  discuss the phenomenological properties of light stops. In section~\ref{FPinLSS} we  review the FP solutions and analyze the conditions under which a light stop may be obtained in the spectrum. In section~\ref{SolFPLSS} we  discuss particular supersymmetry breaking solutions that lead to Light Stops in FP scenarios. Section~\ref{conclusion} will be devoted to our conclusions.

\section{Light Stops}
\label{LightStops}

As discussed in the previous section, the existence of a focus point in the renormalization group evolution of the Higgs mass parameter allows to raise the supersymmetric particle  masses without increasing the low energy value of this parameter. Indeed, the Higgs mass parameter becomes independent of the overall scale of the mass parameters at the messenger scale, and raising the messenger scale values leads to an overall increase of the superpartner masses, without affecting the Higgs mass parameter. However, since the stop masses are affected by similar negative corrections to the ones affecting the Higgs mass parameter, they may be much smaller than the first and second generation squark masses even if they are degenerate at the messenger scale. The presence of light stops is then a generic feature of supersymmetric theories and may lead to interesting phenomenological effects, that may be detected at future experiments.  

Direct searches for top superpartners have started to constrain the light stop parameter space~\cite{Poveda:2014wca}. However, depending on how the stop decays, for stop masses above or of the order of 200~GeV, light stops remain unconstrained.  In particular if the right handed stop is the lightest squark then the searches are not sensitive in a situation when the lightest chargino or neutralino has a mass difference with the stop smaller than about 50 GeV, since in that case the decay of the stop is a three body decay with soft leptons or jets, making the search quite challenging.  Searches for stop production in association with hard jets have started to further restrict this possibility, but currently they loose effectiveness as the stop mass increases beyond 250 GeV.  Limits become stronger when there are also light gluinos, with masses below the TeV scale, but in this article we will not consider this possibility. 

Within the MSSM, the lightest CP-even Higgs mass is determined via radiative corrections by the stop masses. Therefore both stops can not be at the weak scale, since this would lead to a too small lightest CP-even Higgs mass, in contradiction with the experimentally measured Higgs mass value.  Here, we will concentrate on the case in which one of the
two stop supersymmetry breaking mass parameter is much smaller than the other one.  We will choose $m_U$ to be the smallest stop mass parameter, since it is affected by larger quantum corrections and it is unrelated to the sbottom sector, which is strongly constrained by LHC experimental searches. 

\begin{figure}[htb]
\begin{center}
\includegraphics[width=81.9mm]{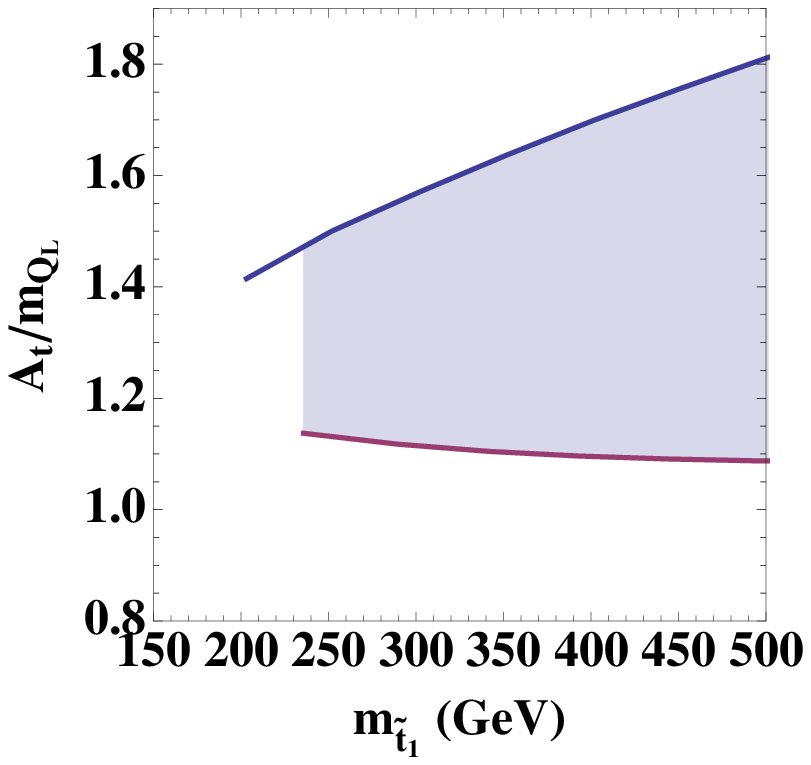}
\includegraphics[width=81.9mm]{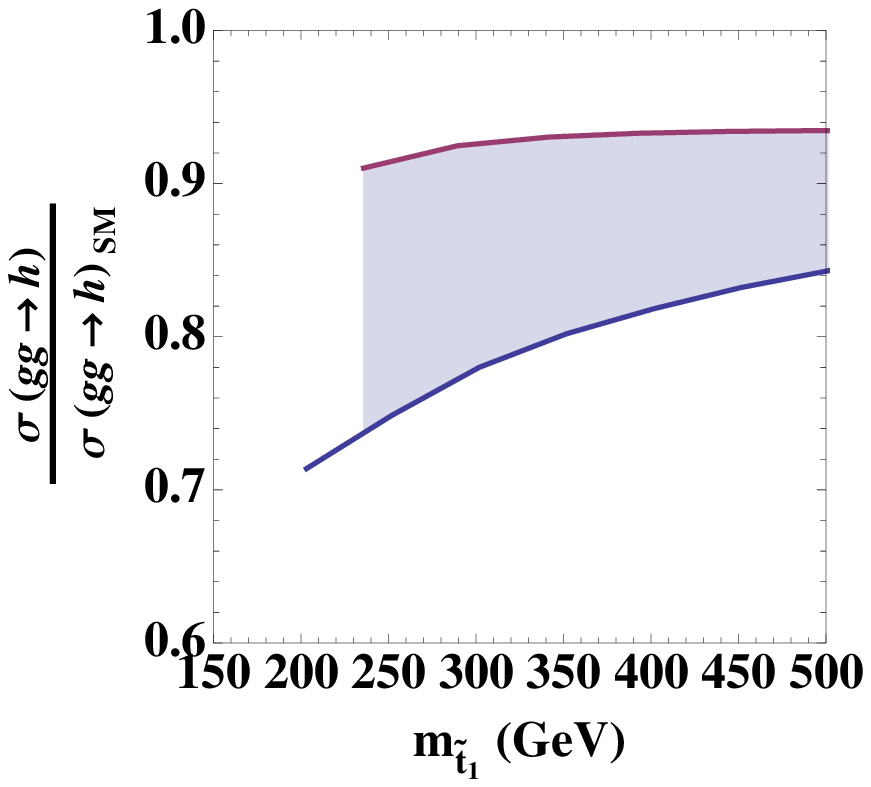}
\end{center}
\caption{\it Left panel :  Values of $A_t/m_{Q_L}$ necessary to reproduce the correct values of the lightest CP-even Higgs mass, $m_h = 125.5 \pm 3$~GeV, as explained in the text, for values of $m_{Q_L} = m_A =2$~TeV and $\tan\beta = 8$. The value of $m_U$ is varied and remains of the order of the lightest stop mass. Right panel : Ratio of the gluon fusion Higgs production rate to the SM value, for the same values of the stop and Higgs mass parameters. Red and blue lines are associated with the smallest and largest allowed values of $A_t$, respectively.}
\label{fig:lightstop1}
\end{figure}

The light stops may affect the Higgs phenomenology by modifying the gluon fusion rate~\cite{Carena:2008mj},\cite{Menon:2009mz},\cite{Curtin:2012aa},\cite{Cohen:2012zza},\cite{Carena:2013iba}. For $m_{Q_L} \gg m_U$, such effects may be suppressed whenever $A_t \simeq m_{Q_L}$.  Indeed, the Yukawa coupling of the light stop to the lightest Higgs is approximately given by
\begin{equation}
g_{h\tilde{t} \tilde t}  \simeq h_t^2 v\left( 1  - \frac{A_t^2}{m_{Q_L}^2} \right).
\end{equation}
For values of $m_{Q_L}$ of the order of a few TeV and $m_U$ of the order of the weak scale, the correct Higgs mass is obtained for values of $A_t \simgt m_{Q_L}$, implying a small or negative effective Yukawa coupling. Interestingly enough a recent analyses~\cite{Camargo-Molina:2014pwa,Blinov:2013fta} show that for these values of the mixing parameter $A_t$, which satisfy the trivial condition $A_t^2<3\, m_{Q_L}^2$~\cite{Frere:1983ag,Kusenko:1996jn},  there are no charge and/or color breaking minima deeper than the electroweak minimum, which guarantees the stability of the electroweak vacuum.  For the smallest values of $|A_t|$, the
Higgs phenomenology is only mildly affected, while for the largest values, the Higgs gluon fusion rate is suppressed, and therefore these values of $A_t$ start to be constrained by current data.  These results are shown in Fig.~\ref{fig:lightstop1}, where we have
used the program CPsuperH~\cite{Lee:2012wa} for the computation of the Higgs masses and we have assumed a theoretical error of about 3~GeV in their determination, namely the countors are consistent with $m_h = (125.5 \pm 3)$~ GeV.  These results differ from the ones presented in Ref.~\cite{Carena:2013iba} since light staus are absent in the spectrum, implying lower values of $BR(h \to \gamma \gamma)$ and slightly larger values of the Higgs mass. Values of $m_{Q_L}$ somewhat larger than 2~TeV would make the Higgs phenomenology variations even milder. 
\begin{figure}[htb]
\begin{center}
\includegraphics[width=81.9mm]{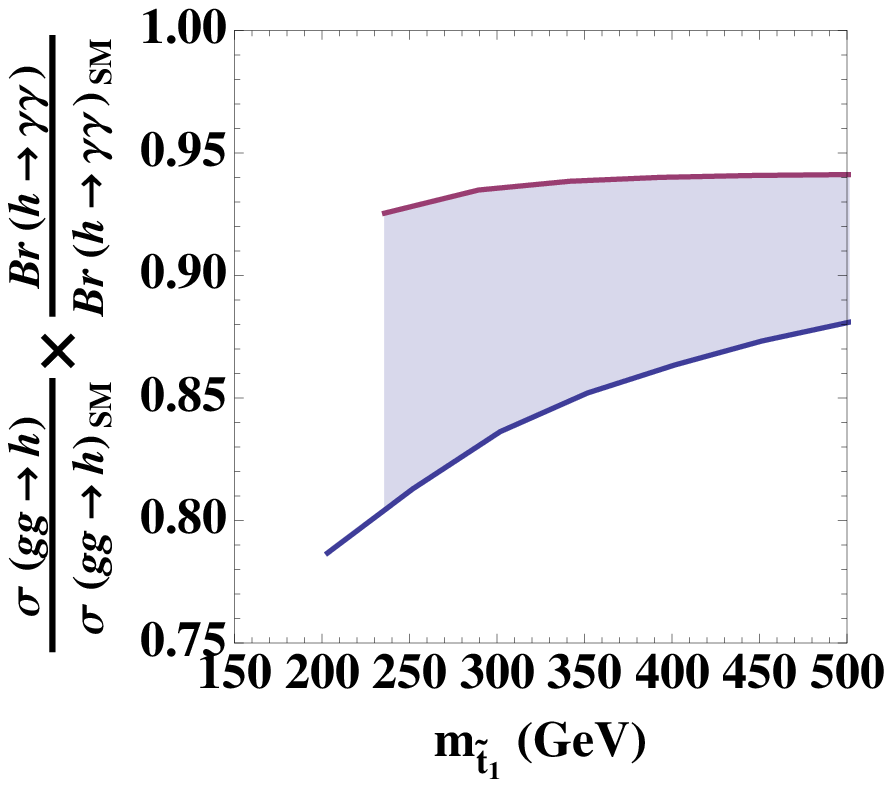}
\includegraphics[width=81.9mm]{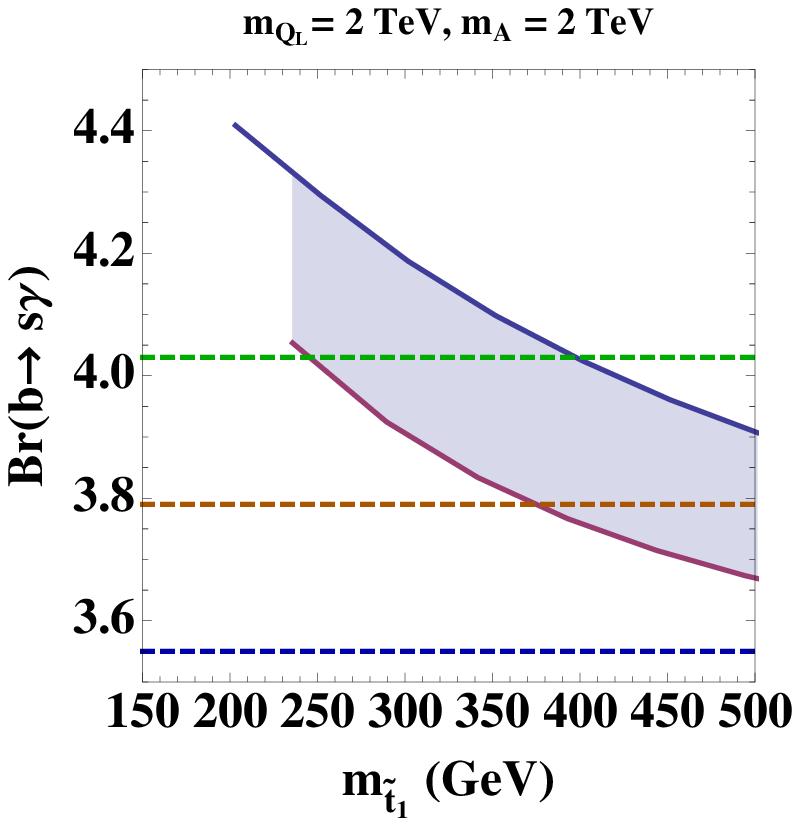}
\end{center}
\caption{\it Left panel: Values of the diphoton production rate for values of $m_{Q_L} = m_A = 2$~TeV and $\tan\beta = 8$.  Red and blue lines are associated with the smallest and largest allowed values of $A_t$, respectively Right panel : Values of $BR(b \to s \gamma)$ obtained for the same values of $m_A$, $m_{Q_L}$ and $A_t$ as in the left panel. The three lines correspond from top to bottom to the $2\sigma$, $\sigma$ and the central value of the experimental data.}
\label{fig:lightstop2}
\end{figure}

The gluon fusion suppression is slightly compensated by an enhancement of the diphoton decay rate, which remains larger than in the SM. In the left panel of Fig.~\ref{fig:lightstop2} we show the total diphoton production rate induced in gluon fusion processes.  Flavor processes may also be affected by the presence of light stops. Indeed, the amplitude of the decay $b \to s \gamma$ is affected not only by the presence of light stops but also by the charged Higgs, and possible small flavor violation processes induced by Yukawa coupling effects in the running from high energies.  Flavor violation processes may also appear at the messenger scale, induced by the supersymmetry breaking mechanism, but we will ignore these effects. The most important contributions come from the charged Higgs and the stop~\cite{Barbieri:1993av}. The charged Higgs induces a correction to the SM amplitude which is of the same sign as the SM one, but suppressed by the charged Higgs mass squared,
\begin{equation}
{\cal{A}}_{H^+}(b \to s \gamma)  \propto  \frac{m_t^2}{m_{H^+}^2}.
\end{equation}
The stop sector, instead, gives a contribution proportional to
\begin{equation}
{\cal{A}}_{\tilde{t}}(b \to s \gamma)  \propto \frac{\mu  A_t  \tan\beta }{m_{Q_L}^2} 
\end{equation}
with a relevant logarithmic dependence on the lightest stop and chargino masses. Large values of $\tan\beta$ would therefore induce large corrections to this process. On the other hand, small values of $\tan\beta$ tend to suppress the Higgs mass. Therefore, moderate values of $\tan\beta \simeq$~10 are preferred in the light stop scenario.

In the right panel of Fig.~\ref{fig:lightstop2} we show the results for $BR(b \to s \gamma)$ for $m_{Q_L} = m_A = 2$~TeV and $\tan\beta = 8$, as computed by CPsuperH in the minimal flavor violating scheme for the given hierarchy between the soft breaking parameters induced by the running from high energies.  We have chosen $\mu = 200$~GeV, but the results vary only slightly provided $\mu$ is not much larger than the weak scale.  We also show in the same plot the experimental central value, as well as  the values allowed at the one and two $\sigma$ confidence level.  As with the Higgs results, flavor processes allow for a light stop, particularly for the smallest values of $A_t$ consistent with the observed Higgs mass. As it happens with the Higgs observables, somewhat larger values of $m_{Q_L}$ will improve the agreement of the predictions of $BR(b \to s \gamma)$ with experimental data.  

The value of the charged Higgs mass may be estimated by considering the condition of electroweak symmetry breaking.  For $m_{H_U}^2 \simeq 0$, obtained in the general FP solution, and moderate or large values of $\tan\beta$, this condition implies a relation between $\tan\beta$, $\mu$ and the non-standard Higgs masses, namely
\begin{equation}
\tan^2\beta \simeq \frac{ m_{H_D}^2  + \mu^2 + M_Z^2/2}{\mu^2 + m_h^2/2},
\label{EoM}
\end{equation}
where the square of the CP-odd mass $m_A^2 \simeq m_{H_D}^2 + 2 \mu^2$. Assuming small values of $\mu$, of order of the weak scale, the value of $m_{H_D}$  is naturally of order $\mu \tan\beta$. These  large values of $m_{H_D}$ are natural in models of non-universal Higgs masses like the ones we will analyze in the next section. For instance, for values of $\tan\beta \simeq 10$, values of the charged Higgs mass of order of a few TeV are obtained. Such large values of the charged Higgs mass lead only to small corrections to $BR(b \to s \gamma)$. 

In the above, we have concentrated on positive values of $A_t$. Negative values of $A_t$ make it somewhat more difficult to obtain the observed Higgs mass and, in addition, lead to values of $BR(b \to s \gamma)$ which are smaller than the 2 $\sigma$ experimental lower bound.  However, somewhat larger 
values of $m_{Q_L}$ would make negative values of $A_t$ consistent with experiment, but we will not explore them in this work.

\section{FP in the LSS}
\label{FPinLSS}

Supersymmetry provides a technical solution to the large naturalness problem (i.e.~why $v\ll M_P$).  However, for large scales of the supersymmetric spectrum $\mathcal Q_0$, consistent with the non-observation of supersymmetric particles at LHC,  an explanation of the little naturalness problem (i.e.~why $v\ll \mathcal Q_0$) is still requireed. One particular solution that alleviates the required amount of fine-tuning is if $\mathcal Q_0$ is the generalized FP solution~\cite{Feng:1999zg} of the running quantity $m_{H_U}^2(\mathcal Q)$, associated with the requirement  that $m_{H_U}^2(\mathcal Q_0)=0$. 

On the other hand, as we have discussed before, the experimental bounds on the lightest (mostly right-handed) stop keep room to have a light enough stop, a scenario dubbed LSS as we explained in the introduction. However a mostly right-handed light stop necessarily requires values of the supersymmetric parameter $m_{U_R}^2$ much smaller than the typical supersymmetric scale $\mathcal Q_0$, thus creating an additional little naturalness problem. A possible solution to this little hierarchy problem is if $\mathcal Q_0$ is additionally a FP of the running  $m_{U_R}^2(\mathcal Q)$, i.e.~if $m_{U_R}^2(\mathcal Q_0) \simeq 0$.  This can be solved by assuming that $m^2_{U_R}(\mathcal Q_0)$ is non zero,  but nevertheless its value (as it is small compared to the rest of soft masses) can be neglected in a first approximation. Since in all the region of parameters
consistent with the observed Higgs mass  and $m_{Q_L}^2(\mathcal Q_0)  \gg m_{U_R}^2(\mathcal Q_0)$,  one obtains $m_{\tilde{t}_1}^2 \simeq m_{U_R}^2(\mathcal Q_0)$, in practice  this amounts to selecting values of $m_{U_R}$ of the order of a few hundred GeV, and 
much smaller than the characteristic mass parameters at the messenger scale. 

In this section we will then deduce the relation on the supersymmetric parameters for $\mathcal Q_0$ to be a double FP characterized by
\be
m_{H_U}^2(\mathcal Q_0) \simeq m_{U_R}^2(\mathcal Q_0) \simeq 0
\label{2FP}
\ee 
In fact we will consider the strict equality in Eq.~(\ref{2FP}) to make general searches on the parameter space and will introduce small realistic masses $m_{U_R}^2(\mathcal Q_0)$ for the particular examples we will present in Sec.~\ref{SolFPLSS}.

We will assume that the MSSM soft breaking terms $(m_{Q_L},\, m_{U_R},\,M_a,\, m_{H_U},\, m_{H_D},\, A_t,\dots)$ are generated at some high-scale $\mathcal M$, where they are communicated to the observable sector by some messenger fields. We use the notation $M_a$ ($a=1,2,3$) for the Majorana masses of the $SU(3)\otimes SU(2)\otimes U(1)$ gauginos while we deserve the notation $Q_L,\, U_R$ to the third generation squark fields~\footnote{Here we are neglecting the Yukawa couplings $h_b$ and $h_\tau$ (as we are never considering too large values of the parameter $\tan\beta$) while the Yukawa couplings of the first two generation are too small and do not play any role.}. The value of $m_Q^2(\mathcal Q)$ for $Q=Q_L,\, U_R,\, H_U$ can be computed on general grounds as
\begin{eqnarray}
m_{Q}^2(\mathcal Q)&=&m_{Q}^2+d_Q \left\{\eta_{Q_L}[\mathcal Q,\mathcal M](m_{Q_L}^2+m_{U_R}^2+m_{H_U}^2)
\phantom{\frac{1^{1^1}}{1^1}}
\right.\nonumber\\
&+&\sum_{a}\left[\eta_{a}[\mathcal Q,\mathcal M]-2 \left(c_{H_U}^a-\frac{c_Q^a}{d_Q}\right)F_a[\mathcal Q,\mathcal M]\right]M_a^2 \nonumber\\
&+&\left.\sum_{a\neq b}\eta_{ab}[\mathcal Q,\mathcal M]M_aM_b+\sum_a \eta_{aA}[\mathcal Q,\mathcal M]M_a A_t+\eta_{A}[\mathcal Q,\mathcal M]A_t^2 \right\}
\label{analytic}
\end{eqnarray}
where the functions $\eta_X$ from the RGE running have been computed semi-analytically in Ref.~\cite{Wagner:1998vd}. In particular by taking the value $\mathcal Q_0=2$ TeV a fit of the functions $\eta_X[\mathcal Q_0,\mathcal M]$ was explicitly done in Ref.~\cite{Delgado:2014vha} in a power series of $\log_{10}(\mathcal M/\textrm{GeV})$. The functions $F_a$ are explicitly given  by 
\be
F_a[\mathcal Q,\mathcal M]=\frac{1}{b_a}\frac{\alpha_a^2(\mathcal M)-\alpha_a^2(\mathcal Q)}{\alpha_a^2(\mathcal M)}=\frac{\alpha_a(\mathcal Q)}{2\pi}\log(\mathcal M/\mathcal Q)\left(2-\frac{b_a\alpha_a(\mathcal Q)}{2\pi}\log(\mathcal M/\mathcal Q)
\right)
\ee
where $b_a=(\frac{33}{5},1,-3)$ and the coefficients are defined, for $Q=(Q_L,U_R,H_U)$, as
\begin{align}
c^3_Q&=(4/3,4/3,0),\quad c_Q^2=(3/4,0,3/4),\quad c_Q^1=(1/60,4/15,3/20) \nonumber\\
d_Q&=(1/3,2/3,1)
\end{align}

\begin{figure}[htb]
\begin{center}
\includegraphics[width=100.9mm]{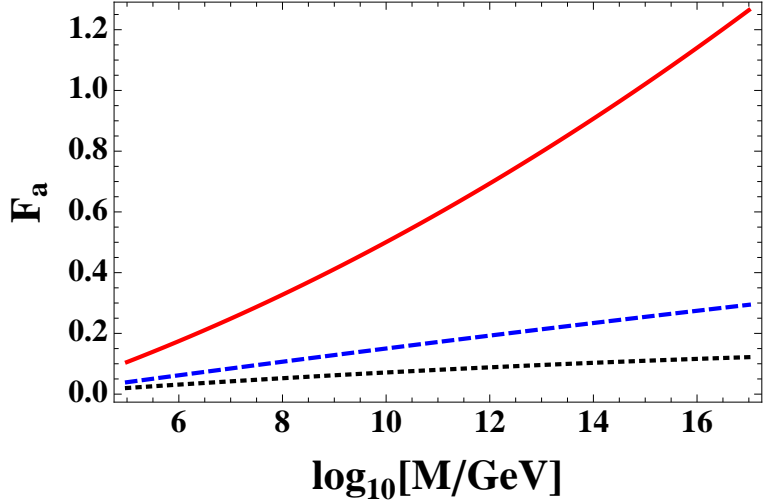}
\end{center}
\caption{\it Plots of $F_3$ (red solid), $F_2$ (blue dashed) and $F_1$ (dotted) as functions of $\log_{10}\mathcal (M/\textrm{GeV})$. }
\label{efes}
\end{figure}

Assuming that $\mathcal Q_0$ is the FP defined by $m_{H_U}^2(\mathcal Q_0)=0$, i.e.
\begin{eqnarray}
0&=&m_{H_U}^2+\eta_{Q_L}[\mathcal Q_0,\mathcal M](m_{Q_L}^2+m_{U_R}^2+m_{H_U}^2)
+\sum_{a}\eta_{a}[\mathcal Q_0,\mathcal M]M_a^2\nonumber\\
&+&\sum_{a\neq b}\eta_{ab}[\mathcal Q_0,\mathcal M]M_aM_b+\sum_a \eta_{aA}[\mathcal Q_0,\mathcal M]M_a A_t+\eta_{A}[\mathcal Q_0,\mathcal M]A_t^2
\label{analytic2}
\end{eqnarray}
one can write the value of $m_{U_R}^2(\mathcal Q_0)$ as given by the expression
\be
m_{U_R}^2(\mathcal Q_0)=m_{U_R}^2-\frac{2}{3}m_{H_U}^2+\sum_a c_a M_a^2F_a[\mathcal Q_0,\mathcal M]
\label{epsilon}
\ee
where $
{\displaystyle (c_1,c_2,c_3)=\left(\frac{1}{3},-1,\frac{8}{3}\right)}$.
The double FP defined in Eq.~(\ref{2FP}) then requires the condition that
\be
m_{U_R}^2=\frac{2}{3}m_{H_U}^2-\sum_a c_a M_a^2F_a[\mathcal Q_0,\mathcal M]
\label{FPLSS}
\ee
The functions $F_a$ then determine when (whether) the LSS FP can be achieved. A plot of them is given in Fig.~\ref{efes}
from where we can see that (by far and depending on the value of $\mathcal M$) the main contribution is that coming from the gluino sector.

At the LSS FP we can give the prediction of $m_{Q_L}^2(\mathcal Q_0)$ as
\be
m_{Q_L}^2(\mathcal Q_0)=m_{Q_L}^2-\frac{1}{2}m_{U_R}^2+\sum_a d_a M^2_a F_a[\mathcal Q_0,\mathcal M]
\ee
where $
{\displaystyle (d_1,d_2,d_3)=\left(-\frac{7}{30},\frac{3}{2},\frac{4}{3}\right) }$, while the prediction of $M_a(\mathcal Q_0)$ is simply given by
\be
M_a(\mathcal Q_0)=\frac{\alpha_a(\mathcal Q_0)}{\alpha_a(\mathcal M)}M_a
\ee

Finally and in a similar way we have computed $A_t(\mathcal Q_0)$ and made a fit as
\be
A_t(\mathcal Q_0)=\sum_a \gamma_a(\mathcal Q_0,\mathcal M) M_a(\mathcal M)+\gamma_A(\mathcal Q_0,\mathcal M) A_t(\mathcal M)
\ee
where we use the fit valid for $\mathcal Q_0=2$ TeV and in the range $\mathcal M\in[10^5,10^{16}]$ GeV
\begin{align}
\gamma_1(\mathcal Q_0,\mathcal M) &= 
 0.0149- 0.0054\,y (\mathcal M)+ 
  0.0001\, y^2(\mathcal M)\nonumber\\
\gamma_2(\mathcal Q_0,\mathcal M) &= 
 0.0924 - 0.0336\, y(\mathcal M) + 
  0.0008\, y^2(\mathcal M)\nonumber\\
\gamma_3(\mathcal Q_0,\mathcal M) &= 
 0.3979- 0.1418\,y(\mathcal M) + 
  0.0021\, y^2(\mathcal M)\nonumber\\
\gamma_A(\mathcal Q_0,\mathcal M) &= 1.2576 - 0.1058\, y(\mathcal M) + 
  0.0030\, y^2(\mathcal M),\
\end{align}
where $y(\mathcal M)\equiv\log(\mathcal M/\textrm{GeV})$.

\section{Particular Scenarios of Supersymmetry Breaking}
\label{SolFPLSS}

In this section we will consider generic cases where the supersymmetry breaking parameters
$(m_{Q_L},\, m_{U_R},\,M_a,\, m_{H_U},\, A_t)$ are such that the double FP equation (\ref{2FP}) is satisfied. The first (trivial) observation is that in CMSSM-type models characterized by $m_{Q_L}=m_{U_R}=m_{H_{U}}\equiv m_0,\quad M_a\equiv m_{1/2}$ there is no scale $\mathcal Q_0$ satisfying Eq.~(\ref{2FP}). In fact from Eq.~(\ref{FPLSS}) and given that the sum $c_aF_a[\mathcal Q_0.\mathcal M]>0$ (as can be seen from Fig.~\ref{efes}) the condition $m_{U_R}^2<\frac{2}{3}m_{H_U}^2$ follows. A simple way out is to give up with the degeneracy of $m_{U_R}$ and/or $m_{H_U}$, as in models dubbed NUHM~\cite{AbdusSalam:2011fc}. This kind of boundary conditions are generic in string constructions~\cite{Brignole:1993dj} where the soft breaking mass of a scalar field is fixed by its modular weight. Non-universal gaugino masses may also be considered~\cite{Horton:2009ed,Agashe:1999ct}. In particular large values of $M_2$, that has the negative coefficient in $F_2$, Eq.~(\ref{FPLSS}), can induce positive corrections to the Higgs mass parameter without affecting the right-handed stop mass parameter, making it possible to fulfill the double FP condition even for universal scalar masses. However, since very large values of $M_2$ would be required for this to happen, and string constructions provide, at tree level, universal gaugino masses,  we will only concentrate in the following on the case of non-universal scalar masses.

\subsection{Non-Universal Higgs Masses}
\begin{figure}[htb]
\begin{center}
\includegraphics[width=81.9mm]{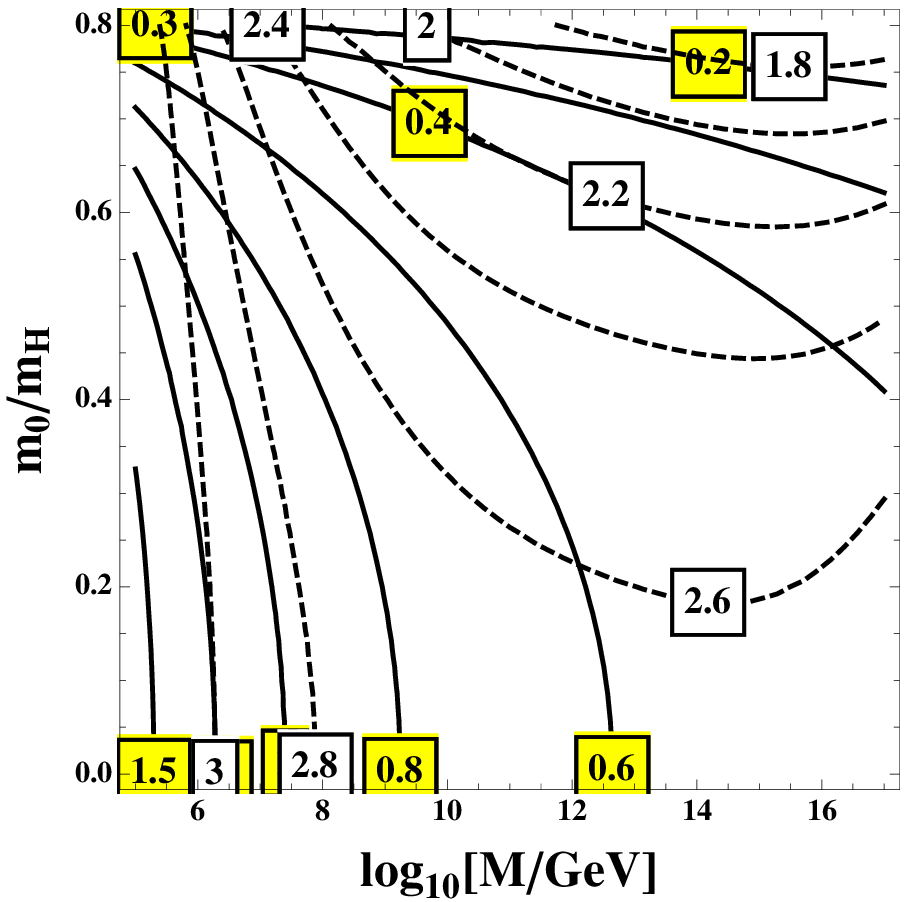}
\includegraphics[width=81.9mm]{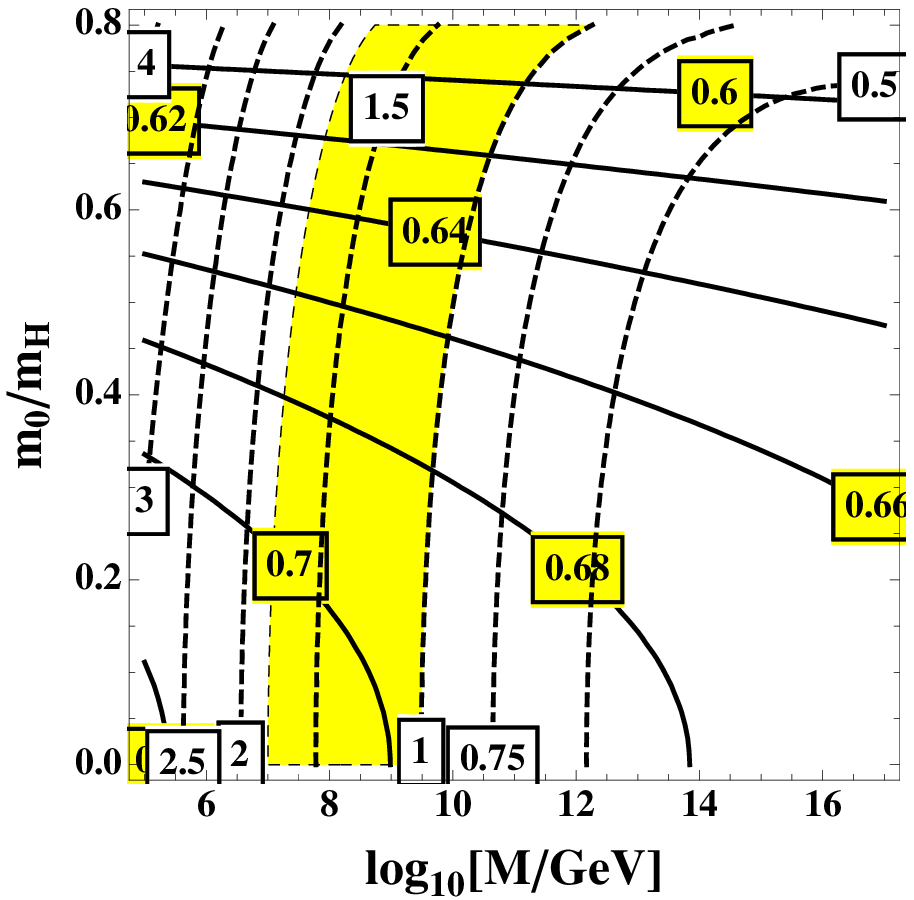}
\end{center}
\caption{\it Left panel: Solid (dashed) lines are contour lines of $m_{1/2}/m_H$ ($A_{t}/m_H$ for $A_t>0$) in the plane $(\log_{10}[\mathcal M/\textrm{GeV}],m_0/m_H)$. Right panel: Solid (dashed) lines are contour lines of $m_{Q_L}(\mathcal Q_0)/m_H$ ($A_{t}(\mathcal Q_0)/m_{Q_L}(\mathcal Q_0)$  for $A_t(\mathcal Q_0)>0$ ) in the plane $(\log_{10}[\mathcal M/\textrm{GeV}],m_0/m_H)$. Shadowed region corresponds to $1\lesssim A_t(\mathcal Q_0)/m_{Q_L}(\mathcal Q_0)\lesssim 1.8$.}
\label{NUHM1}
\end{figure}

\begin{figure}[htb]
\begin{center}
\includegraphics[width=81.9mm]{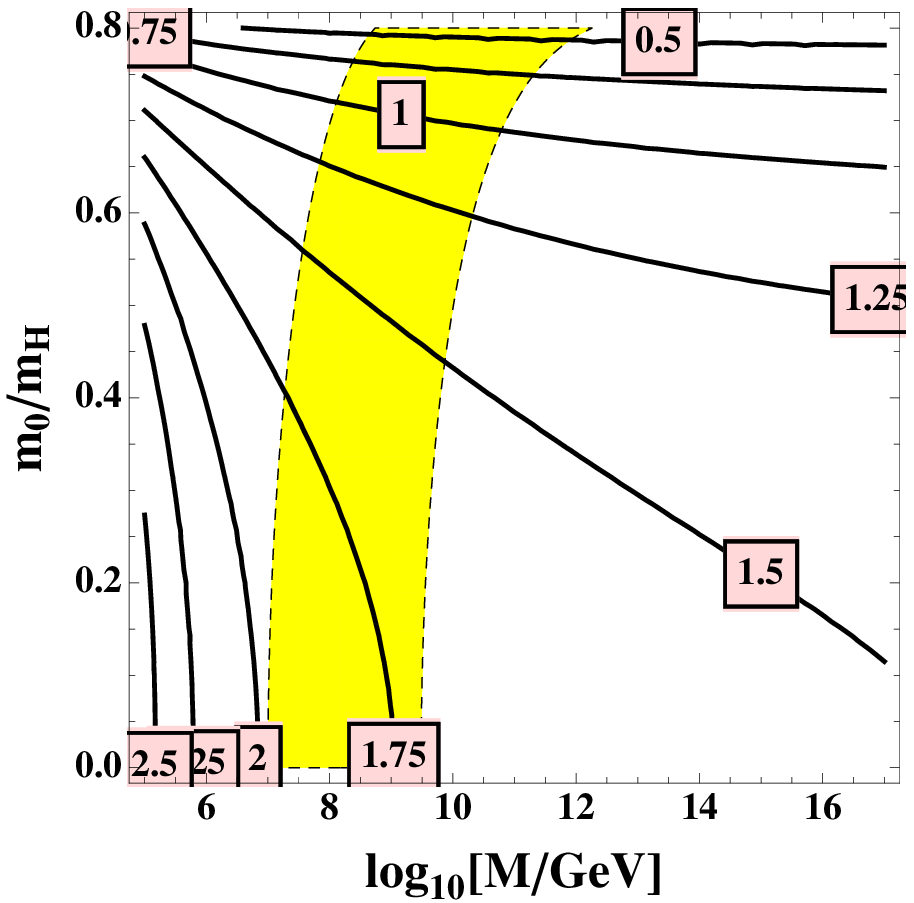}
\includegraphics[width=81.9mm]{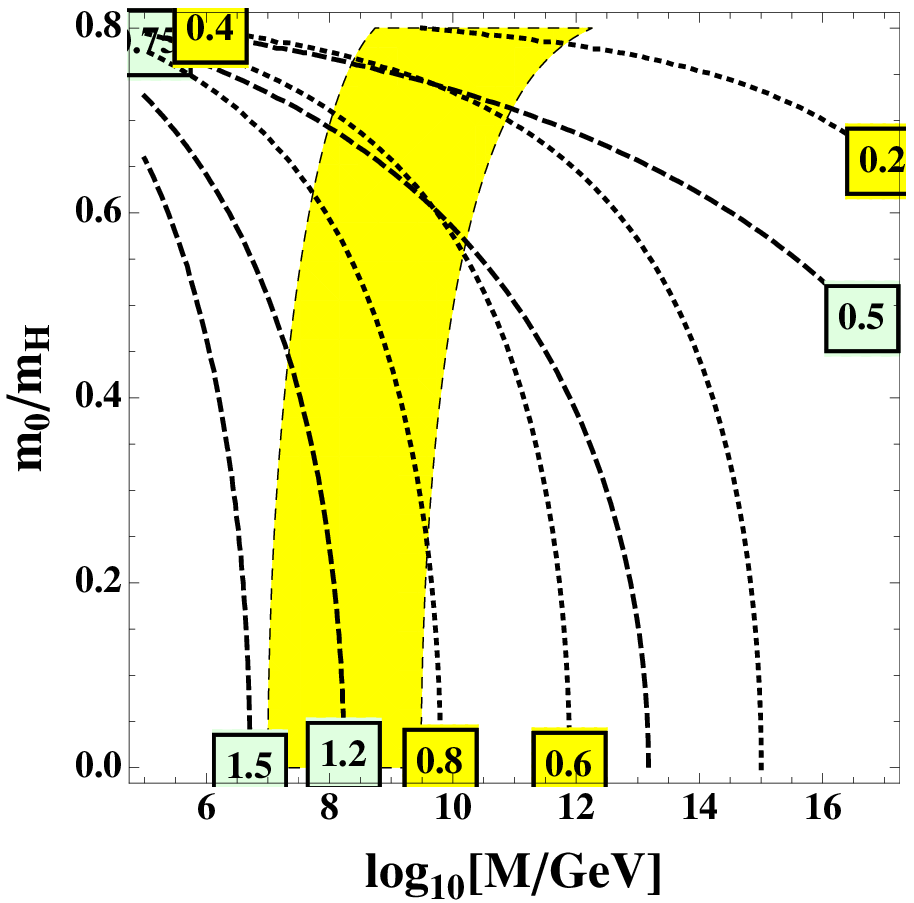}
\end{center}
\caption{\it Left panel: Contour lines of $M_{3}(\mathcal Q_0)/m_{Q_L}(\mathcal Q_0)$ in the plane $(\log_{10}[\mathcal M/\textrm{GeV}],m_0/m_H)$. Right panel: Contour lines of $M_{2}(\mathcal Q_0)/m_{Q_L}(\mathcal Q_0)$ (dashed lines) and $M_{1}(\mathcal Q_0)/m_{Q_L}(\mathcal Q_0)$ (dotted lines) in the plane $(\log_{10}[\mathcal M/\textrm{GeV}],m_0/m_H)$. Shadowed region corresponds to $1\lesssim A_t(\mathcal Q_0)/m_{Q_L}(\mathcal Q_0)\lesssim 1.8$.}
\label{NUHM2}
\end{figure}

We will consider the supersymmetric parameters where at the messenger scale the Higgs sector has a different mass from the one of the squark-slepton sector, namely the four independent parameters at the scale $\mathcal M$ are
\be
m_{Q_L}=m_{U_R}\equiv m_0,\quad A_t,\quad  m_{H_U}=m_{H_D}\equiv m_H,\quad M_a\equiv m_{1/2}
\label{HCMSSM}
\ee
and will impose the double FP condition for $\mathcal Q_0$ as in Eq.~(\ref{2FP}).  The results are shown in Figs.~\ref{NUHM1} and \ref{NUHM2}.

In the left panel of Fig.~\ref{NUHM1} 
we show contour lines of constant $m_{1/2}/m_H$ (solid lines) and $A_t/m_H$ in the plane $\left[\log(\mathcal M/\textrm{GeV}),m_0/m_H\right]$ where we only consider the region where $A_t>0$ in agreement with the results from Sec.~\ref{LightStops}. In fact we can see that there is an upper bound $m_0^2/m_H^2\leq 2/3$ which is reached for massless gauginos $m_{1/2}/m_H=0$. A particular realization of a model with $m_0^2 \simeq 2/3\ m_H^2$ and $m_{1/2}\ll m_H$, was proposed in Ref.~\cite{oai:arXiv.org:1201.5164}, where the heavy scalars are generated by gauge mediation of an extra $U(1)_\chi\otimes U(1)_F$ group spontaneousy broken at high scales, while the masses of light gauginos are generated by gravity mediation.

More important for phenomenological purposes are the values of supersymmetric parameters at the scale $\mathcal Q_0$. 
In the right panel of Fig.~\ref{NUHM1} we show contour lines of $m_{Q_L}(\mathcal Q_0)/m_H$ (solid lines) and of $A_t(\mathcal Q_0)/m_{Q_L}(\mathcal Q_0)$ (dashed lines). In particular we can see that the (shadowed) region $1\lesssim A_t(\mathcal Q_0)/m_{Q_L}(\mathcal Q_0)\lesssim 1.8$ allowed by the Higgs mass determination is consistent with scales $10^7 \textrm{ GeV }\lesssim \mathcal M\lesssim 10^{12}\textrm{ GeV}$. In this region we see that there is a window of values of $m_{Q_L}(\mathcal Q_0)$ as $0.6\lesssim m_{Q_L}(\mathcal Q_0)/m_H\lesssim 0.7$.
As for the values of $M_a(\mathcal Q_0)/m_{Q_L}(\mathcal Q_0)$ they are plotted in the left panel  of Fig.~\ref{NUHM2} for $M_3(\mathcal Q_0)$ (solid lines), and in the right panel  of Fig.~\ref{NUHM2} for $M_2(\mathcal Q_0)$ (dashed lines) and for $M_1(\mathcal Q_0)$ (dotted lines). In particular in the selected region the gluino mass is in the window $0.5\lesssim M_3(\mathcal Q_0)/m_{Q_L}(\mathcal Q_0)\lesssim 2$. In all cases the band allowed by the Higgs mass is superimposed in the contour plots of supersymmetric parameters at the low scale $\mathcal Q_0$.

As a simple estimate and to get a feeling of the order of magnitude of the involved parameters we will fix $m_{Q_L}(\mathcal Q_0)\simeq 2$ TeV which is achieved in the selected region for
\be
2.9 \textrm{ TeV}\lesssim m_H\lesssim 3.3\textrm{ TeV},\quad 
0<m_0\lesssim 2.5\textrm{ TeV}
\ee
and then implies that 
\begin{eqnarray}
2\textrm{ TeV}&\lesssim A_t(\mathcal Q_0)&\lesssim 3.6\textrm{ TeV}\nonumber\\
1.5\textrm{ TeV}&\lesssim M_3(\mathcal Q_0)&\lesssim 4\textrm{ TeV}\nonumber\\
700\textrm{ GeV}&\lesssim M_2(\mathcal Q_0)&\lesssim 3\textrm{ TeV}\nonumber\\
500\textrm{ GeV}&\lesssim M_1(\mathcal Q_0)&\lesssim 2.5\textrm{ TeV}
\end{eqnarray}
where we have constrained $m_0\lesssim 0.75\  m_H$  to not have too light gluino masses.
A particular case~\footnote{In this and in the next examples we have considered realistic values $200\textrm{ GeV}\lesssim m_{U_R}(\mathcal Q_0)\lesssim 500$ GeV, as we will specify.} is given in Tab.~\ref{tablaNUHM} where the messenger scale $\mathcal M=10^{10}$ GeV is selected and the input parameters at the messenger scale are
\begin{table}[htb]
\begin{center}
\begin{tabular}{||c|c|c||c|c|c|c|c|c||}
 \hline\hline
$\frac{m_{0}}{m_H}$&$\frac{A_t}{m_H}$ & $\frac{m_{1/2}}{m_H}$&$\frac{m_{Q_L}(\mathcal Q_0)}{m_H}$& $\frac{m_{U_R}(\mathcal Q_0)}{m_{Q_L}(\mathcal Q_0)}$& $\frac{A_t(\mathcal Q_0)}{m_{Q_L}(\mathcal Q_0)}$&$\frac{M_3(\mathcal Q_0)}{m_{Q_L}(\mathcal Q_0)}$&$\frac{M_2(\mathcal Q_0)}{m_{Q_L}(\mathcal Q_0)}$&$\frac{M_1(\mathcal Q_0)}{m_{Q_L}(\mathcal Q_0)}$
 \\ \hline
 0.6&2.35&0.52& 0.65&0.22 &1.03 &1.27&0.74&0.59 \\
\hline\hline
  \end{tabular}
 \caption{\it A particular set of parameters at $\mathcal M=10^{10}$ GeV (left set) and low $\mathcal Q_0$ energy (right set) from Figs.~\ref{NUHM1} and~\ref{NUHM2}.}
 \label{tablaNUHM}
 \end{center}
 \end{table}
given in the left side of Tab.~\ref{tablaNUHM} while the output parameters at the scale $\mathcal Q_0$ are given on its right side. We have also used as an input parameter, in Eq.~(\ref{epsilon}), $m_{U_R}^2(\mathcal Q_0)=0.02\,m_H^2$. If we fix $m_{Q_L}(\mathcal Q_0)=2$ TeV it corresponds to $m_{U_R}(\mathcal Q_0)\simeq 440$ GeV, $m_H\simeq 3.1$ TeV and $m_0\simeq 1.9$ TeV, $m_{1/2}\simeq 1.6$ TeV, $A_t\simeq 7.5$ TeV. On the other hand the gluino and  electroweak gaugino low energy masses $M_a(\mathcal Q_0)$ are $(2.52, 1.48,1.18)$ TeV, for $a=3,2,1$, respectively.

\subsection{Non-Universal Scalar Masses}
As we have seen in the previous section from the double FP condition, in view of the current experimental constraints on the Higgs and gluino masses,  the transmission of supersymmetry breaking at the GUT scale $\mathcal M_{GUT}\simeq 2\times 10^{16}$ GeV cannot be achieved for non-universal Higgs masses and universal gaugino masses. A way out is to give up the universality of the squark masses, a generic situation which appears on soft breaking terms coming from superstring theories~\cite{Brignole:1993dj}. 
\begin{figure}[htb]
\begin{center}
\includegraphics[width=81.9mm]{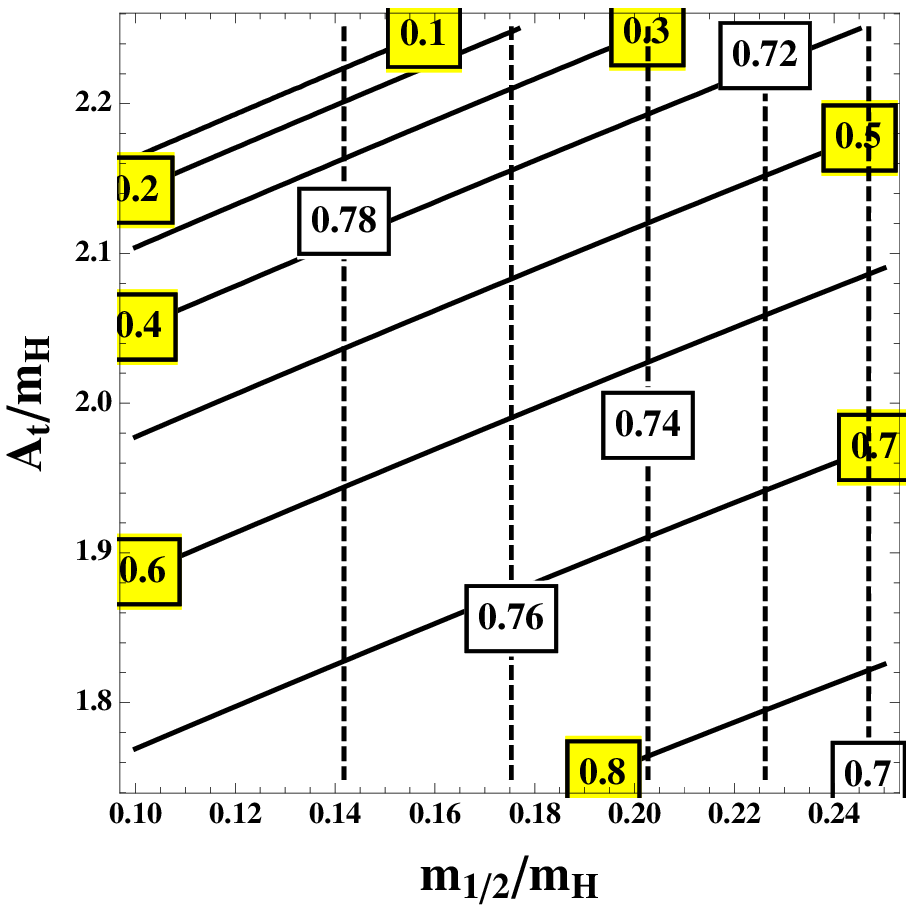}
\includegraphics[width=81.9mm]{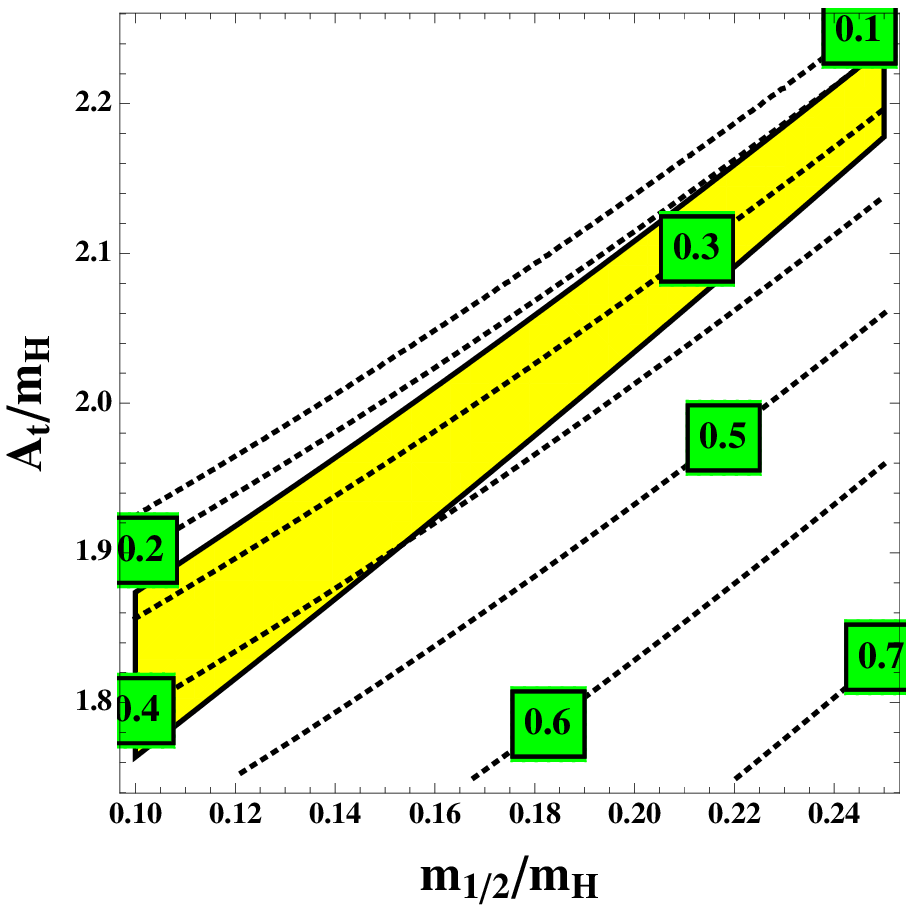}
\end{center}
\caption{\it Contour lines for $\mathcal M_{GUT}=2\times 10^{16}$ GeV. Left panel: Solid (dashed) lines are contour lines of $m_{Q_L}/m_H$ ($m_{U_R}/m_H$) in the plane $(m_{1/2}/m_H,A_t/m_H)$. Right panel: Solid lines are contour lines for $A_t(\mathcal Q_0)/m_{Q_L}(\mathcal Q_0)=$1 (lower line) and 1.8 (upper line). Shadowed region is allowed by the Higgs mass. Dotted lines are contour lines for $m_{Q_L}(\mathcal Q_0)/m_H$ in the plane $(m_{1/2}/m_H,A_t/m_H)$.}
\label{NUSM1}
\end{figure}
Therefore in this section we will consider soft breaking terms characterized by the five independent parameters defined at the unification scale $\mathcal M_{GUT}$
\be
m_{Q_L},\quad m_{U_R},\quad A_t,\quad m_{H_U}\equiv m_H,\quad M_a\equiv m_{1/2}
\label{NUSM}
\ee
on which we will impose the double FP condition (\ref{FPLSS}). The results are shown in Figs.~\ref{NUSM1} and~\ref{NUSM2}. This kind of boundary condition could potentially produce a non-zero Fayet-Iliopoulos (FI) term in the RGE evolutions of soft masses. As its impact on the present calculation is proportional to $g_1^2$ and  therefore tiny,  we are going to assume in the remaining of this subsection that the rest of scalar masses are such that the FI cancels.

In the left panel of Fig.~\ref{NUSM1} we show contour lines of $m_{Q_L}/m_H$ (solid lines) and $m_{U_R}/m_H$ (dashed lines) in the plane $(m_{1/2}/m_H,A_t/m_H)$ and, as we did in the previous section, we have selected positive values $A_t>0$, in agreement with  phenomenological requirements on the LSS as we showed in Sec.~\ref{LightStops}. Contour lines for $m_{Q_L}(\mathcal Q_0)/m_H$ are shown in the right panel of Fig.~\ref{NUSM1} (dotted lines) along with the contours of $A_t(\mathcal Q_0)/m_{Q_L}(\mathcal Q_0)=1$ (lower solid line) and  $A_t(\mathcal Q_0)/m_{Q_L}(\mathcal Q_0)=1.8$ (upper solid line) such that the available region which can accommodate the experimental value of the Higgs mass is the shadowed (yellow) region.

\begin{figure}[htb]
\begin{center}
\includegraphics[width=81.9mm]{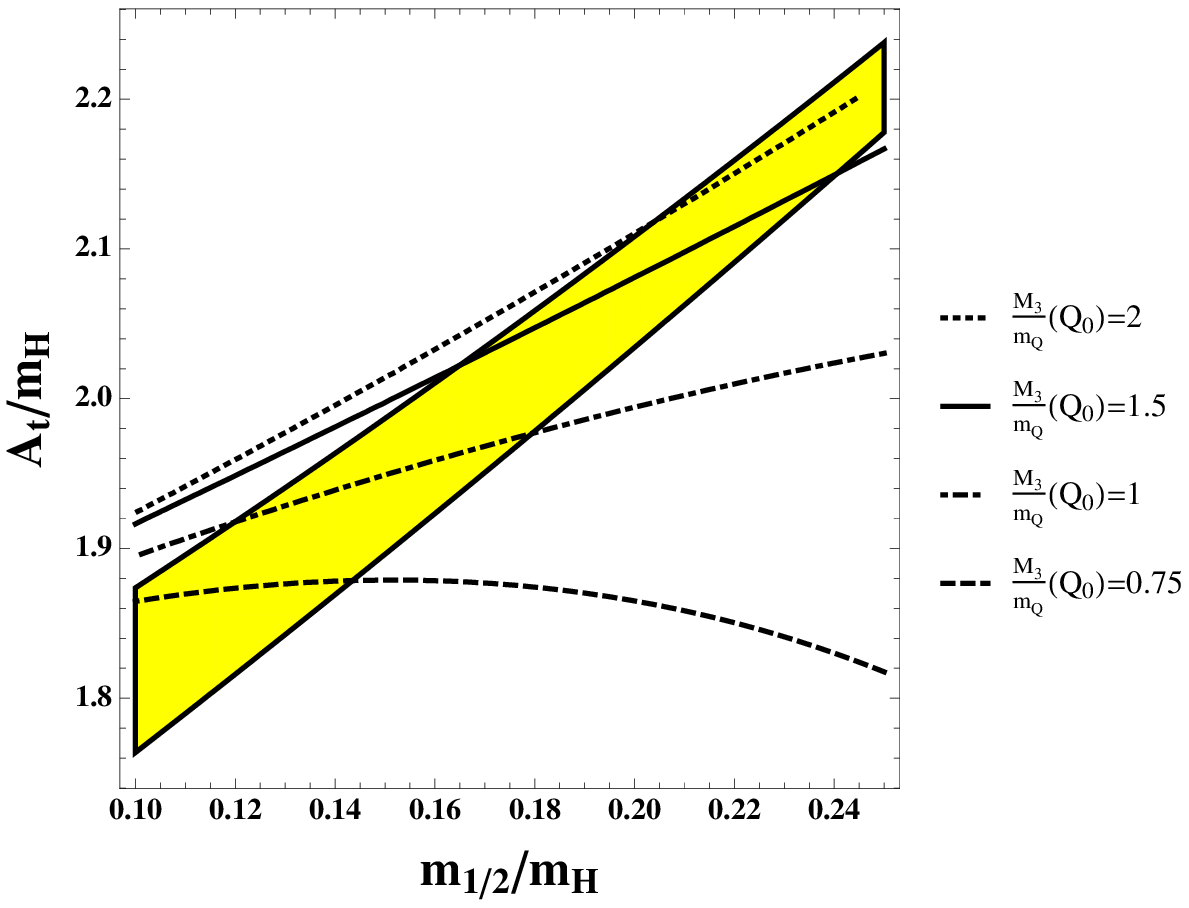}
\includegraphics[width=81.9mm]{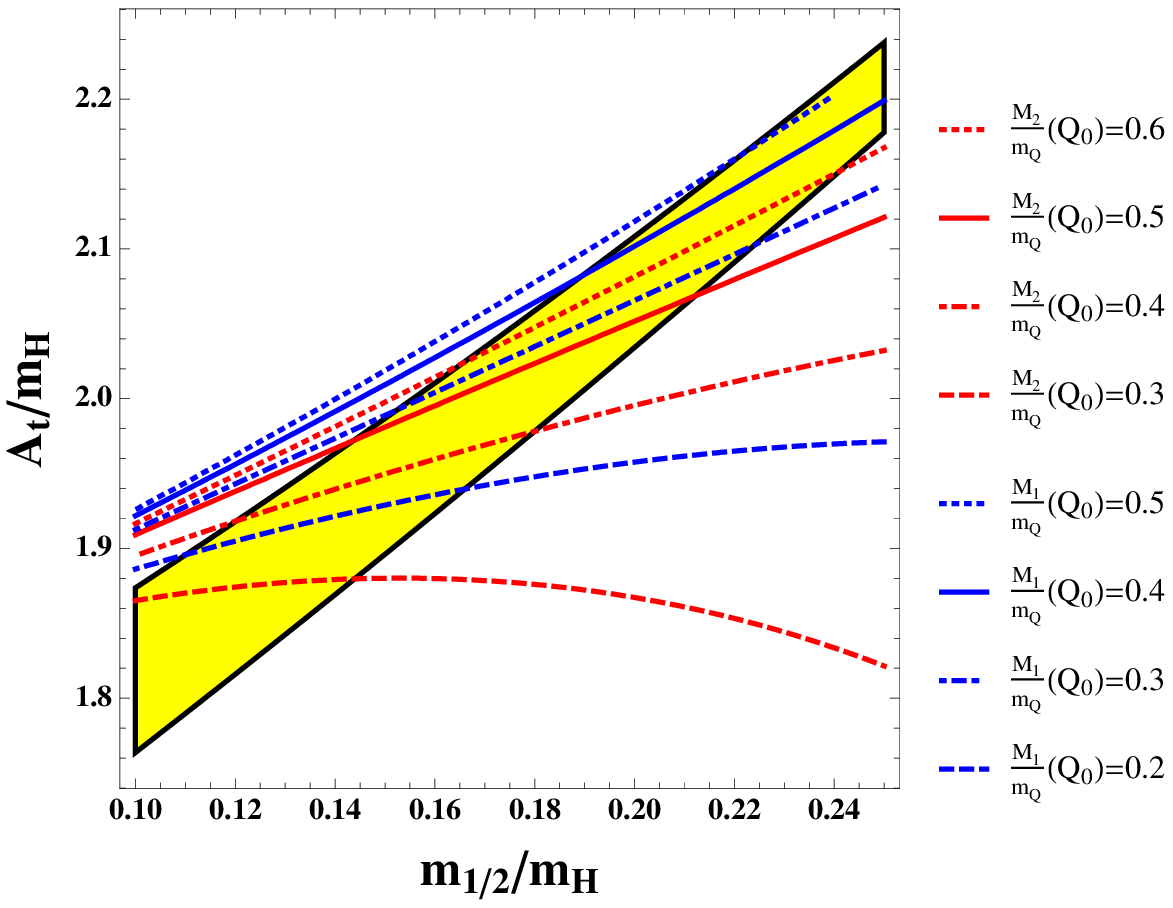}
\end{center}
\caption{\it Contour lines for $\mathcal M_{GUT}=2\times 10^{16}$ GeV. Left panel: Contour lines of gluino mass ratios $M_3(\mathcal Q_0)/m_{Q_L}(\mathcal Q_0)$ in the plane $(m_{1/2}/m_H,A_t/m_H)$. Right panel: Contour lines of electroweak gaugino mass ratios $M_2(\mathcal Q_0)/m_{Q_L}(\mathcal Q_0)$ and $M_1(\mathcal Q_0)/m_{Q_L}(\mathcal Q_0)$ in the plane $(m_{1/2}/m_H,A_t/m_H)$. The allowed region by the Higgs mass is superimposed.}
\label{NUSM2}
\end{figure}
Contour plots for $M_a(\mathcal Q_0)/m_{Q_L}(\mathcal Q_0)$ are exhibited in Fig.~\ref{NUSM2}. In particular in the left panel contour lines of the gluino mass ratio $M_3(\mathcal Q_0)/m_{Q_L}(\mathcal Q_0)$ (solid lines) are presented while the electroweak gaugino mass ratios are exhibited in the right panel for $M_2(\mathcal Q_0)/m_{Q_L}(\mathcal Q_0)$ (dashed lines) and for $M_1(\mathcal Q_0)/m_{Q_L}(\mathcal Q_0)$ (dotted lines). In all cases the region where the theory is consistent with the Higgs mass is superimposed with the other contour lines.

 \begin{table}[htb]
\begin{center}
\begin{tabular}{||c|c|c|c||c|c|c|c|c|c||}
 \hline\hline
$\frac{m_{1/2}}{m_H}$&$\frac{A_t}{m_H}$ & $\frac{m_{Q_L}}{m_H}$&$\frac{m_{U_R}}{m_H}$&$\frac{m_{Q_L}(\mathcal Q_0)}{m_H}$&$\frac{m_{U_R}(\mathcal Q_0)}{m_{Q_L}(\mathcal Q_0)}$&$\frac{A_t(\mathcal Q_0)}{m_{Q_L}(\mathcal Q_0)}$&$\frac{M_3(\mathcal Q_0)}{m_{Q_L}(\mathcal Q_0)}$&$\frac{M_2(\mathcal Q_0)}{m_{Q_L}(\mathcal Q_0)}$&$\frac{M_1(\mathcal Q_0)}{m_{Q_L}(\mathcal Q_0)}$
 \\ \hline
 0.22&2.1&0.55&0.73& 0.33&0.22&1.08 &1.42&0.57&0.32 \\
\hline\hline
  \end{tabular}
 \caption{\it One particular set of parameters at high ($\mathcal M_{GUT}$) and low $(\mathcal Q_0)$ energy from Figs.~\ref{NUSM1} and~\ref{NUSM2} leading to heavy electroweak gauginos.}
 \label{tablaNUSM1}
 \end{center}
 \end{table}
\begin{table}[htb]
\begin{center}
\begin{tabular}{||c|c|c|c||c|c|c|c|c|c||}
 \hline\hline
$\frac{m_{1/2}}{m_H}$&$\frac{A_t}{m_H}$ & $\frac{m_{Q_L}}{m_H}$&$\frac{m_{U_R}}{m_H}$&$\frac{m_{Q_L}(\mathcal Q_0)}{m_H}$&$\frac{m_{U_R}(\mathcal Q_0)}{m_{Q_L}(\mathcal Q_0)}$&$\frac{A_t(\mathcal Q_0)}{m_{Q_L}(\mathcal Q_0)}$&$\frac{M_3(\mathcal Q_0)}{m_{Q_L}(\mathcal Q_0)}$&$\frac{M_2(\mathcal Q_0)}{m_{Q_L}(\mathcal Q_0)}$&$\frac{M_1(\mathcal Q_0)}{m_{Q_L}(\mathcal Q_0)}$
 \\ \hline
 0.13&1.85&0.66&0.79& 0.40&0.18&1.05 &0.70&0.28&0.15 \\
\hline\hline
  \end{tabular}
 \caption{\it Another particular set of parameters at high ($\mathcal M_{GUT}$) and low $(\mathcal Q_0)$ energy from Figs.~\ref{NUSM1} and~\ref{NUSM2} leading to lighter electroweak gauginos.}
 \label{tablaNUSM2}
 \end{center}
 \end{table}
Two typical examples are provided in Tabs.~\ref{tablaNUSM1} and~\ref{tablaNUSM2} where we present sets of input parameters in units of $m_H$, at the scale $\mathcal M_{GUT}$ (left side of tables) and the corresponding output parameters at the scale $\mathcal Q_0$ (right side of tables). We have fixed here $m_{U_R}^2(\mathcal Q_0)=0.005\, m_H^2$. The model in Tab.~\ref{tablaNUSM1} corresponds to heavy gauginos while that in Tab.~\ref{tablaNUSM2} corresponds to lighter gauginos. In fact fixing $m_{Q_L}(\mathcal Q_0)=2$ TeV amounts then to $m_H\simeq 6$ TeV, and $m_{U_R}(\mathcal Q_0)\simeq 440$ GeV for the case of Tab.~\ref{tablaNUSM1} and $m_H\simeq 5$ TeV, and $m_{U_R}(\mathcal Q_0)\simeq 360$ GeV for the case of Tab.~\ref{tablaNUSM2}. The gaugino masses at the low scale $M_{a}(\mathcal Q_0)$ are given by $(2.84,1.14,0.64)$ TeV, for $a=3,2,1$ respectively, for the case of Tab.~\ref{tablaNUSM1}, and $(1.4,0.56,0.30)$ TeV, for $a=3,2,1$ respectively, for the case of Tab.~\ref{tablaNUSM2}. 
 
 \subsection{Electroweak Symmetry Breaking and the Charged Higgs Mass}
 
 The value of $m_{H_D}$ at the low scale $\mathcal Q_0$ has to satisfy the EoM, Eq.~(\ref{EoM}), which leads to $m_{H_D}(\mathcal Q_0)\simeq \tan\beta\, \mu(\mathcal Q_0)$. As the value of $m_{H_D}$ at the high scale $\mathcal M$ does decouple from the one-loop calculation of the double FP condition, and as it renormalizes by a little amount because of the bottom Yukawa and electroweak gauge couplings we can assume the relation $m_{H_D}\simeq \tan\beta\, \mu$  which can then be  used to fix $m_{H_D}$ or $ \mu$.  
 
Imposing the cancellation of the hypercharge FI D-term contribution, the soft supersymmetry breaking parameter $m_{H_D}^2$ 
 may be determined as a function of the other scalar mass parameters
 \begin{equation}
m_{H_D}^2 = 3  \left(m_{Q_L}^2 - 2 \ m_{U_R}^2 + m_{D_R}^2 - m_{L_L}^2 + m_{E_R}^2 \right) + m_{H_U}^2 ,
 \end{equation}
 where the factor 3 comes from the number of generations and we have assumed flavor universality. For simplicity, we 
 will assume that the mass difference in the slepton sector is zero or small compared to the ones appearing in the squark and Higgs
 sector.  In such a case, $m_{H_D}$ is simply determined as a function of $m_{H_U}$ and the squark mass parameters. 
 
As an example, let us  assume that $m_{D_R}^2 = m_{U_R}^2$.  Then,
 \begin{equation}
\frac{m_{H_D}^2}{m_{H_U}^2} =  3  \left( \frac{m_{Q_L}^2}{m_{H_U}^2} - \frac{m_{U_R}^2}{m_{H_U}^2} \right)  + 1.
 \label{mhdmhu}
 \end{equation}
Since in the LSS at the messenger scale $m_{Q_L} < m_{U_R}$, one obtains a reduction of $m_{H_D}$ with respect to $m_{H_U}$.  For instance, if
one takes the values of the parameters displayed in the upper right corner of the yellow shaded region in Fig.~\ref{NUSM1}, $m_{Q_L} = 0.45 m_H$
and $m_{U_R} = 0.7 m_H$, we obtain  
\begin{equation}
\frac{m_{H_D}^2}{m_{H_U}^2} \simeq 0.1.
\end{equation}
Since $m_{H}/m_{Q_L}(\mathcal Q_0) \simeq 5$ then $m_{H}$ is of order 10~TeV and $m_{H_D}$ and the charged Higgs mass would be of order 3~TeV.
Eq.~(\ref{mhdmhu}) leads to similar values of the charged Higgs mass obtained throughout the yellow shaded region in Fig.~\ref{NUSM1}.
 
 If, instead, we assumed that $m_{D_R} = m_{Q_L}$, then the values of the charged Higgs mass become smaller, and consistent solutions, without
 tachyons,  may only   be obtained for small values of the gaugino masses in Fig.~6.  These examples simply show that for reasonable boundary conditions of the scalar mass parameters, the condition of cancellation of the FI term tends to induce values of the charged Higgs mass that are of order $m_{Q_L}(\mathcal Q_0)$, significantly smaller than $m_H$.  This, in turn, is consistent with the condition of electroweak symmetry breaking for small values of $\mu$ and moderate values of $\tan\beta$, and as we show in section~\ref{LightStops}, may lead to agreement with collider and flavor constraints in the LSS.  
 
 \subsection{Lightest Neutralino, Dark Matter and Stop Searches}
 
 Quite generally, in the generalized focus point scenario, the value of $\mu$ is small and then the lightest neutralino has a significant Higgsino component.  Light Higgsinos lead to a thermal relic density that is far smaller than the one observed experimentally and therefore either non-thermal production or other sources of Dark-Matter are required in this case.   
 
 If Higgsinos are the only particle lighter than the light stop, then the stop would tend to decay into a bottom quark and a charged Higgsino. The charged Higgsino, in turn, would decay into a neutral Higgsino and soft quarks and leptons, which would be difficult to observe at the LHC. Therefore, at the LHC stop decays would lead to bottom quarks and missing energy and therefore standard sbottom searches (with sbottoms  decaying into bottom quark and missing energy) may be used to constrain this scenario~\cite{Aad:2013ija}.  Stop masses within a few tens of GeV of the neutralino mass would remain unconstrained in this case. 
 
In certain cases, the gaugino masses are small and then a thermal Dark Matter may be obtained with a well tempered neutralino condition~\cite{ArkaniHamed:2006mb}. For instance, Table~\ref{tablaNUSM2} present a case in which the Bino mass is of order 300~GeV and naturally of order $\mu$. In general, $M_1$  has only a small impact on the renormalization group evolution of the scalar mass parameters and therefore the value of $M_1$ may be lowered without affecting the general properties of the FP solution.  Direct Dark Matter detection mediated by the CP-even Higgs bosons put  constraints on this scenario, which become weaker for negative values of $M_1$~\cite{Ellis:2000jd,Baer:2006te,Cheung:2012qy,Huang:2014xua}.  On the other hand, in such a case the mass difference between charginos and neutralinos become larger  than in the light Higgsino scenario and searches for stops proceed in the standard $b + W$+ Missing Energy channel that we discussed in Sec.~\ref{LightStops}. 
 
\section{Conclusions}
\label{conclusion}

In this article we have discussed the possibility of obtaining a double focus point, light stop scenario,  in which both the right-handed stop mass parameter $m_{U_R}^2$ and the Higgs mass parameter $m_{H_U}^2$ become independent of the generic supersymmetric particle mass scale. 
We have required the gluino to be heavy and a stop mixing parameter $A_t$ such that the right Higgs mass is obtained without affecting in a drastic way the Higgs phenomenology.  For this to happen, the low energy mixing parameter should be of the order or somewhat larger than the heaviest stop mass, namely $A_t (\mathcal Q_0) \simgt m_{Q_L}(\mathcal Q_0)$.

The requirement of obtaining a light stop within the focus point scenario demands particular correlations of the scalar and gaugino mass parameters at the messenger scale. In particular, this cannot be achieved in the case of universal scalar and gaugino masses. We have shown, however, that a double focus fixed point may be achieved in the case of non-universal Higgs mass parameters.  Even in such a case, the LSS demands  a messenger scale significantly smaller than  the GUT scale $\mathcal M_{GUT} \simeq 2 \times 10^{16}$~GeV.  

In view of these constraints, we have studied the conditions under which a light stop may be obtained with a messenger scale of the order of the GUT scale. Concentrating on the case of universal gaugino masses, we have determined the specific scalar mass parameter correlations that lead to the presence of a double focus point within the LSS.  For this to happen, for not too heavy gluino masses,  the scalar mass parameters at the GUT scale must remain of the same order, but they must fulfill specific correlations, namely,  $ m_{U_R}^2/m_{H_U}^2 \simlt 0.65$, and $m_{Q_L}^2/m_{H_U}^2 \simlt 0.5$, while the mixing parameter, $1.5 \ m_{H_U} \simlt A_t \simlt 2.5  \ m_{H_U}$.   Finally, the universal gaugino masses should acquire moderate values, $m_{1/2} \simlt 0.3 \ m_{H_U}$.  

We have also shown that the condition of electroweak symmetry breaking demands the CP-odd Higgs boson mass $m_A \simeq \mu \tan\beta$.  For low energy values of $m_{Q_L}(\mathcal Q_0)$ of a few TeV,  moderate values of $\tan\beta$ are required in order to obtain the proper Higgs mass  while keeping agreement with flavor physics constraints, in particular the measured value of the $BR(b \to s \gamma)$.  For the specific non-universal scalar masses analyzed in this article, the value of $m_{H_D}^2$ at the messenger scale must be smaller than $m_{H_U}^2$, what can be obtained in simple supersymmetry breaking scenarios.

\section*{Acknowledgments}
The work of AD was partially supported by the National Science Foundation under grant  PHY-1215979. The work of MQ was supported in part by the European Commission under the ERC Advanced Grant BSMOXFORD 228169, by the Spanish Consolider-Ingenio 2010 Programme CPAN (CSD2007-00042), by CICYT-FEDER-FPA2011-25948 and by Secretaria d�Universitats i Recerca del Departament d'Economia i Coneixement de la Generalitat de Catalunya under Grant number 2014 SGR 1450. The work of CW at ANL is supported in part by the U.S. Department of Energy under Contract No. DE-AC02-06CH11357.

 \end{document}